\documentclass{aa}
\usepackage{graphics}
\newcommand{\Halp}{H$_{\alpha} $ }
\newcommand{\Hbet}{H$_{\beta} $ }
\newcommand{\Hgam}{H$_{\gamma}{ }$ }

\newcommand{\HeII} {HeII(4686\AA) }

\newcommand{\am}{AM Her }

\begin{document}                                          
\thesaurus{02.09.1, 02.13.1, 13.25.5, 08.02.3, 08.23.3, 08.09.2: AM Herculis}
\title{Magnetic field and unstable accretion during AM Herculis low states.
\thanks{Based on data collected at SAO (Russia), CAO (Crimea), BAO (Bulgaria)
and OHP (France)}}

\author{J.M. Bonnet-Bidaud\inst{1}
\and M. Mouchet\inst{2,3} 
\and N.M. Shakhovskoy\inst{4}
\and T.A. Somova\inst{5}
\and N.N. Somov\inst{5}
\and I.L. Andronov\inst{6}
\and D.~de Martino\inst{7}
\and S.V. Kolesnikov\inst{6}
\and Z. Kraicheva\inst{8}}

\institute
{Service d'Astrophysique, DSM/DAPNIA/SAp, CE Saclay, 
F-91191 Gif sur Yvette Cedex, France
\and
DAEC, Observatoire de Paris Section de Meudon, F-92195 Meudon Cedex, France
\and
Universit\'{e} Denis Diderot, Place Jussieu, F-75005 Paris, France
\and
Crimean Astrophysical Observatory, Nauchnyi, Crimea, 334413, Ukraine
\and
Special Astrophysical Observatory, Nizhnij Arkhyz,
Karachaj-Cherkess Republik, 357147, Russia
\and
Department of Astronomy , Odessa St University, Shevchenko Park, Odessa, 270014, Ukraina
\and
Osservatorio Astronomico di Capodimonte, I-80131 Napoli, Italy
\and 
Institute of Astronomy, Bulgarian Academy of Sciences, 72 Tsarigradsko Shouse Blvd., 
1784 Sofia, Bulgaria
}
\offprints{J.M. Bonnet-Bidaud}
\mail{bobi@discovery.saclay.cea.fr}
\date{Received date: 29 September 1999; accepted date: 23 December 1999}
\titlerunning{Magnetic field and unstable accretion in AM Her}
\authorrunning{Bonnet-Bidaud et al.}
\maketitle

 \begin{abstract}
A study of AM Her low states in September 1990 and 1991 and June-July 1997 
is reported from  a coordinated campaign  
with observations obtained at the Haute-Provence observatory,
 at the 6-m telescope of the Special
Astrophysical Observatory and at the 2.6m and 1.25m telescopes of the 
Crimean observatory.
Spectra obtained at different dates when the source was 
in low states at a comparable V magnitude, show  
the presence of strong Zeeman absorption features and marked changes
in emission lines with  a day-to-day reappearance of the HeII (4686\AA) 
emission lines in 1991.
Despite this variability, the magnetic field inferred from the fitting of 
the absorption spectrum with Zeeman hydrogen splitting, is remarkably
constant with a best value of (12.5$\pm$0.5)MG.
Detailed analysis of the UBVRI light curves shows the presence 
of repetitive moderate amplitude ($\sim$ 0.3-0.5 mag) flares
 predominantly red in colour. 
These flares are attributed to small accretion events and are
compared to the large ($\sim$ 2 mag.) blue flare reported by  Shakhovskoy et al. (1993). 
We suggest that the general flaring activity observed during the low states 
is generated by accretion events.
The different characteristics of the flares (colour and polarization) 
are the results of different shock geometries depending on the net mass accretion 
flux.

\end{abstract} 

\section{Introduction}
AM Herculis (4U1814+49) is the well-known prototype of the "polar" systems,
a subclass of cataclysmic variables in which a highly magnetized 
(10$^{7}$G) white dwarf in a close binary system accretes matter from 
a low-mass companion (see Cropper 1990, Chanmugam 1992 for a review).
From the long term optical monitoring collected and kindly made available
to us by the AAVSO (J. Mattei, private communication), it is now 
well known to oscillate irregularly between two different states of optical
brightness, 
a high state at (V $\sim$12.5) corresponding to high accretion 
rate and a low state (V $\sim$15) during which the accretion 
luminosity is significantly reduced so that photospheric emission of the 
two stars becomes visible in the infrared for the companion and in the UV 
and the optical for the white dwarf.
The low state  reveals in particular a complicated optical spectrum where
strong absorption features due to the Zeeman splitting of the Balmer lines 
produced in the high surface magnetic field of the white dwarf are clearly
visible, allowing the direct measurement of the field (Schmidt et al. 1981, 
Latham et al. 1981, Young et al. 1981). 
During low states, the UV emission is also found consistent with  
white dwarf atmosphere models with T=(2-2.5) 10$^{4}$K and typical size
(6-8) 10$^{6}$m (Heise \& Verbunt 1988, Gänsicke et al. 1995, Silber et al. 1996).\\
The question of whether or not the accretion ceases during low states  
is still open.
UV data indicate that the polar caps are still
substantially heated and the few low states observed in the  X-rays reveal 
the presence of residual accretion. A weak X-ray modulation due to  the occultation
of the main accreting pole is detected (Fabbiano 1982, de Martino 
et al. 1998). At such low rates, the accretion onto the white dwarf
is highly unstable and eventually switches-off as recently 
observed by the BeppoSAX satellite (de Martino et al. 1998).
Strong (30\%) quasi-periodic optical oscillations with
periods near 5 minutes have been observed during a decline to a low state
 and interpreted as an accretion instability arising
close to the capture radius (Bonnet-Bidaud et al. 1991). 
A spectacular sharp rising ($\sim$1hr) 
flare of $\sim$2 mag. was also detected during a 1992 low state, tentatively
associated with a stellar flare from the red dwarf companion
(Shakhovskoy et al. 1993). 
We present here data obtained during different low states of AM Her
in 1990, 1991 and 1997 which show that the source presents very different 
characteristics and variability. 
The 1997 data were obtained to supplement contemporaneous BeppoSAX X-ray 
observations (de Martino et al. 1998).

\section{Observations}

\begin{table*}
\caption{Journal of observations.}
\medskip
\begin{tabular}{lllllrr}
\multicolumn{1}{c}{\em Date} & \multicolumn{1}{c}{\em Telescope} &
  \multicolumn{1}{c}{\em Observations} & \multicolumn{1}{c}{\em Range} &
  \multicolumn{1}{c}{\em Resolution} & \multicolumn{1}{c}{\em HJD start} & 
 \multicolumn{1}{c}{\em Exp(min)} \\   \hline
1990 Sep. 17& SAO 6m   & spectroscopy      & 3800-5800\AA&  4\AA      &  8152.2342&  144\\
1991 Sep. 04& SAO 6m   & spectroscopy      & 3950-4950\AA&  2\AA      &  8504.2340&  102\\ 
      "     & SAO 6m   & photometry        &         B   &  0.1s      &  8504.2405&   83\\
      "     & CAO 1.25m& photom./polarim.  &    UBVRI    & 23.1s/180s &  8504.2497&  224\\
1991 Sep. 05& SAO 6m   & spectroscopy& 3950-4950\AA& 2\AA &  8505.2330&    85\\
      "     & SAO 6m   &  photometry &    B        & 0.1s &  8505.2488&   61  \\
      "     & CAO 1.25m& photom./polarim.  &    UBVRI    & 23.1s/180s &  8505.2543&  243\\
1997 June. 30& SAO 6m   & spectroscopy      & 4050-5070\AA&  2\AA     & 10630.4583&  196\\ 
1997 July 01& OHP 1.93m& spectroscopy      & 3750-7250\AA& 13\AA      & 10631.3647&  90\\
            & CAO 2.6m & photom./polarim.  & R           & 4s/128s    & 10631.3329& 244\\
            & CAO 1.25m& photom./polarim.  & UBVRI       & 20s/192s   & 10631.3711& 182\\
1997 July 03& BAO 0.6m & photometry        & V           & 60s        & 10633.4066& 225\\
1997 July 04& BAO 0.6m & photometry        & V           & 60s        & 10634.4074& 242\\
1997 July 29& OHP 1.93m& spectroscopy      & 3750-7250\AA& 13\AA      & 10659.3967& 180\\            
1997 July 30& CAO 1.25m& photom./polarim.  &    UBVRI    & 23.1s/180s & 10660.3166&  177\\
1997 July 31& OHP 1.93m& spectroscopy      & 3750-7250\AA& 13\AA      & 10660.5985&  45\\            
      "     & CAO 1.25m& photom./polarim.  &    UBVRI    & 23.1s/180s & 10661.3513&   82\\
\hline
\end{tabular}
\end{table*}

\subsection{ Photometry and polarimetry}

Photometric and polarimetric observations were conducted 
at the 1.25m AZT-11 telescope of the Crimean Astrophysical 
Observatory (CAO), 
equipped with the double-beam chopping polarimeter of the Helsinski University 
(Korhonen et al.\,1984). On 1991 Sept 4 and 5 and 1997 July 1 and 30,
UBVRI data were collected  during  $\sim$ 3-4 h intervals overlapping spectroscopic observations 
described below. The UBVRI photometric data were recorded automatically with 
a 23.1s resolution and the polarimeter was used in circular 
polarization mode with a resolution of 3 min in the same bands. 
However a statistically significant signal was only received    
in circular polarization in the R and I bands and only the corresponding 
data are discussed here. 
The full description of the polarimeter and the method
of the observations are presented in Berdyugin \& Shakhovskoy (1993). 
On 1997 July 1, data were also obtained at the CAO 2.6m Shajn telescope, 
using a one-channel polarimeter with a fast rotating achromatic quarter-wave 
plate, in the wide R-band (0.50-0.75nm) with a 4s integration time.

\begin{table*}
\caption{Magnitudes in quiescence.}
\medskip
\begin{tabular}{llllllllllll}
\multicolumn{1}{c}{\em Date} & \multicolumn{1}{c}{\em } &
  \multicolumn{2}{c}{\em U} & \multicolumn{2}{c}{\em B} &
  \multicolumn{2}{c}{\em V} & \multicolumn{2}{c}{\em R} & 
  \multicolumn{2}{c}{\em I}  \\  \hline
1992 Aug 29(*) & Q92 & 14.90 &        & 15.58&          &15.40   &          & 14.79&          & 13.56&       \\
1991 Sep.  4   & Q4  & 14.72 & (0.11) &15.52 &  (0.09)  & 15.11  &  (0.17)  & 14.49 & (0.19)  & 13.41 &  (0.18) \\
1991 Sep 5     & Q5  &  14.57& (0.13) &15.38 &  (0.08)  &  15.08 & (0.20)   &  14.31&  (0.21)   &13.32  & (0.20)   \\
1997 July 1    & Q1  &  14.82& (0.12) &15.55 &  (0.11)  &  15.34 & (0.24)   &  14.70&  (0.08)   &13.60  & (0.11)   \\
1997 July 30   & Q30 &  14.81& (0.12) &15.53 &  (0.09)  &  15.16 & (0.22)   &  14.61&  (0.19)   &13.44  & (0.22)   \\
  \hline
  \multicolumn{5}{l} {(*) from Shakhovskoy et al. (1993)}\\
  \multicolumn{6}{l} {Numbers in parentheses are a measure of the dispersion}\\
\end{tabular}
\end{table*}

On 1991 Sept 4 and 5, photometric data were acquired at 
the 6-m BTA telescope of the Special 
Astrophysical Observatory (SAO) (Nizhnij Arkhyz, Russia). 
The observations were carried out at the Nasmyth focus of the telescope, 
simultaneously with the spectroscopy, using 50\% of the incoming flux split 
by a dichroic plate to the NEF photometer (Vikuliev et al.\, 1991). 
A light curve through a Johnson B-filter with a 12 arcsec aperture was 
recorded with a 0.1s resolution during the observations. UBVR measurements 
were also performed at the beginning and end of the observations to calibrate 
the brightness level of the source.
On 1997 July 3 and 4, complementary photometric data were also obtained
at the Belogradchik Astronomical Observatory BAO (Bulgaria), 
using an ST-8 CCD camera attached to a 60cm telescope.\\
All UBVR magnitudes were obtained from differential measurements, 
using the star D (m$_{V}$ = 13.1) in the field as a comparison 
(Liller 1977, Andronov \& Korotin, 1982).\\

\begin{figure*}
\resizebox{\hsize}{!}{\includegraphics{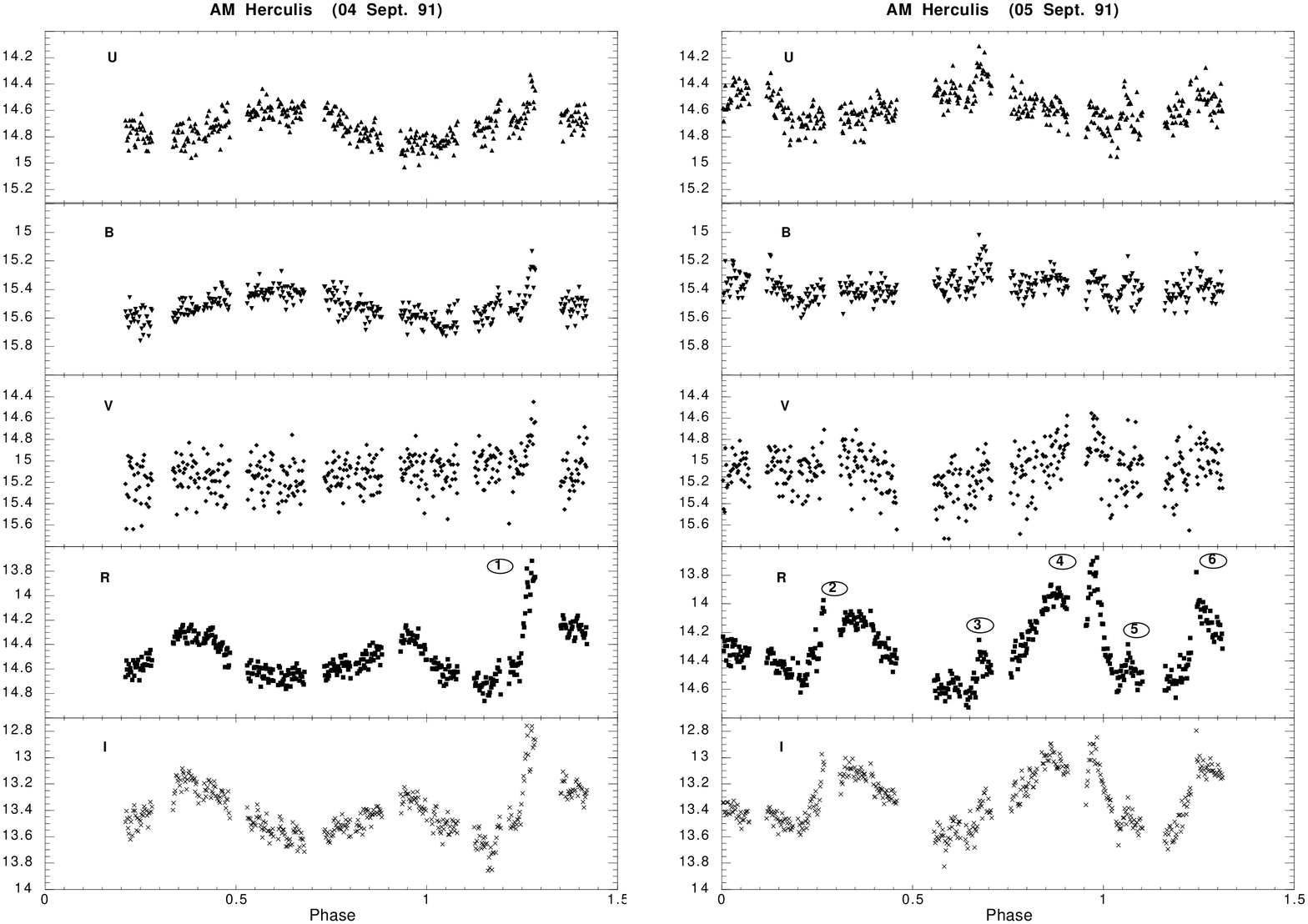}}
%\picplace{12.0 cm}
%\psfig{figure=AMfig1a.eps,width=1a8.cm}
%\scaledpicture 132mm by 99mm (Figure1 scaled 700)
\caption[ ]{UBVRI light curves of the \am low state in 1991. 
(left Sept. 4, right Sept. 5).
Clear flaring activity is seen. 
The strongest flares have been marked by numbers }
\end{figure*}

\begin{figure*}
\resizebox{\hsize}{!}{\includegraphics{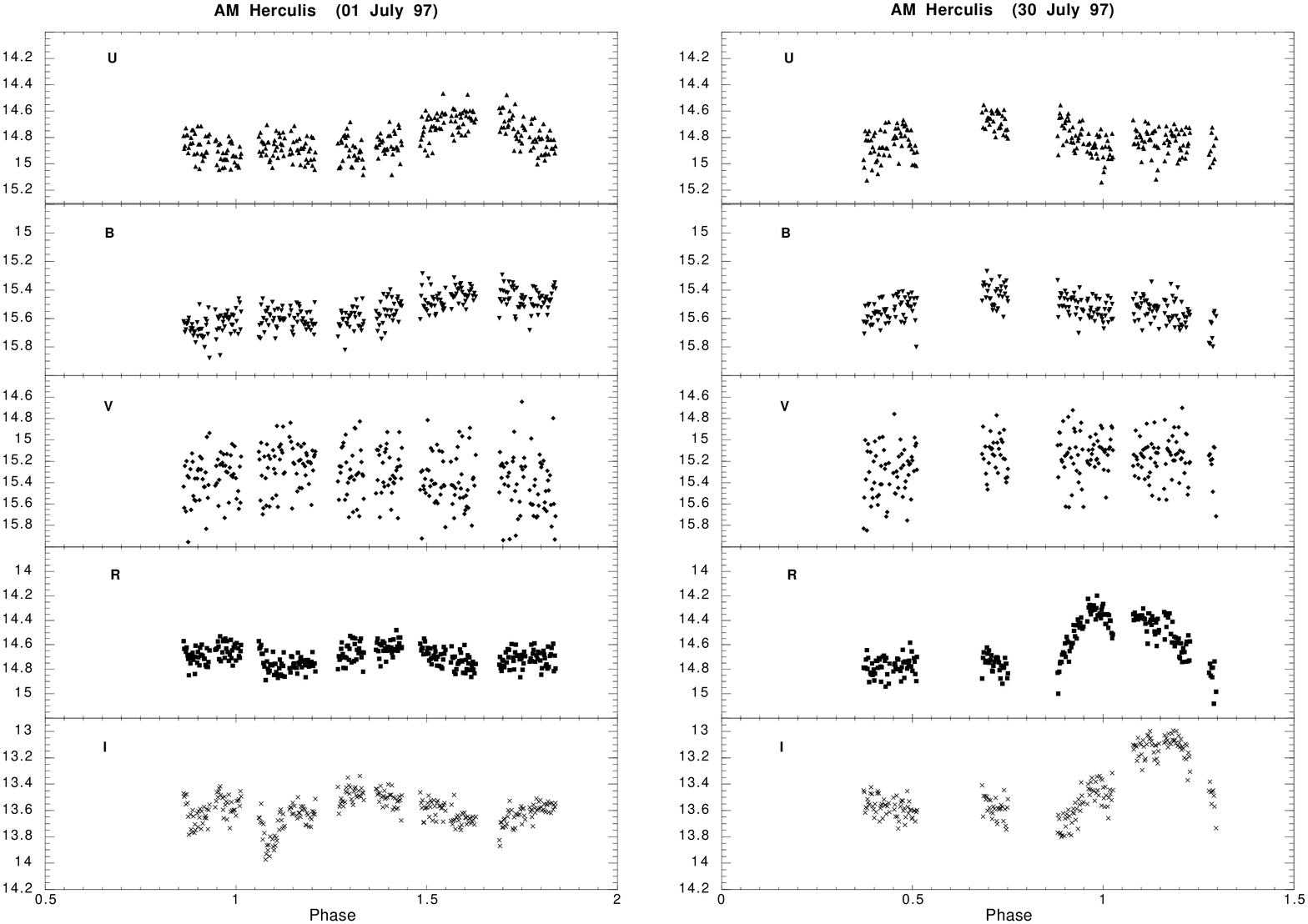}}
%\picplace{12.0 cm}
%\psfig{figure=GRFIG1.PS,width=8.cm}
%\scaledpicture 132mm by 99mm (Figure1 scaled 700)
\caption[ ]{UBVRI light curves of the \am low state in 1997
(left July 1, right July 30). 
No clear modulation is seen except for the flux increase 
at the end of July 30 }
\end{figure*}

\subsection{Spectroscopy}
Spectroscopic data were collected during \am low states 
in 1990, 1991 and 1997 using the SP-124 spectrograph of the 
6-m BTA telescope (Ioannisiani et al. 1982). 
A television scanner with two lines of 1024 channels is used to record the 
sky and source spectra simultaneously in a photon-counting mode 
(see Somova et al. 1982, for a detailed description of the instrumentation). 
A 2.5 arcsec aperture was set, adapted to the seeing and the 
spectrograph was equipped with a grating yielding a resolution of 4\AA\,, 2\AA\, and 
2\AA\,, respectively in 1990, 1991 and 1997.
Data were reduced using the SIPRAN software developed at SAO (Somov 1986).
Part of the 1990 spectroscopic data was already preliminarily discussed 
in Bonnet-Bidaud et al.(1992).\\
On 1997, spectroscopic observations were also obtained at the 
1.93m telescope of the Haute-Provence Observatory (OHP),
equipped with the Carelec spectrograph (Lemaitre et al. 1990).\\
The log of the observations is presented in Table 1 with the Heliocentric 
Julian Dates (HJD) corresponding to the start of the exposures.
In the following, the orbital/rotational phases have been computed according 
to the ephemeris derived by S. Tapia and quoted in Heise \&
Verbunt (1988), where HJD($\phi=0$)= 2443014.76614(4)+0.128927041(5)E, 
 $\phi=0$ corresponding to the maximum of linear polarization.

\section{Analysis and results}
\subsection{ Photometry and polarimetry}    
Figures 1 and 2 show the UBVRI light curves covering more than one 
orbital cycle in the different observations. 
The source is seen at a mean V magnitude of $\sim$ 15.1
and $\sim$ 15.2, respectively in 1991 and 1997, 
consistent with the level previously reported during low states of the star
(Szkody et al. 1982, Bailey et al. 1988). 
Table 2 gives the mean magnitude and dispersion in the different bands 
and at the different epochs.

In 1991, the shape of the modulation is typical of a low state 
with a broad hump
in U and B around phase 0.6, antiphased with a minimum in R and I. 
A secondary minimum in R and I is also visible around phase 0.2.
The V light curve is flat but with a greater dispersion.
Superimposed on the smooth modulation, clear rapid (5-10  min.) flux 
enhancements are also seen, particularly visible in R and I bands. 
Characteristics of these flares (marked with numbers on Figure 1)
are analysed below.
In 1997, the source is less active, though at a similar brightness level
with no evidence of flares.
For all observing nights, the modulation is nearly absent in all bands and 
no strong flux variations 
are seen, except for a dip around phase 0.1 in I on July 1 and a $\sim$0.3 mag. 
broad increase in the R-band near phase 0. on July 30.
BAO light curves obtained on July 3 and 4 also show no modulation.\\
Figure 3a and b shows the corresponding circular polarization in the R and I bands.
A significant polarization is seen in 1991, with a mean level 
in the R / I bands of -3.9\% /-2.6\% and -4.5\% /-4.0\% on Sept 4 
and 5 respectively, indicating a significant residual accretion. 
Negative polarization reaches a maximum of up to -12\% around phase 0.4 
and 0.7 and a minimum near 0\% around phase 0.1, comparable to what usually
observed during normal high states of the source (Bailey \& Axon 1981).    
The polarization, attributed to cyclotron radiation, is usually 
restricted to the infrared bands during low state observations 
 and shifted to higher frequencies only during high states
(Bailey et al. 1988). It is usually absent in R and I bands during low 
states (Shakhovskoy et al. 1993, Silber et al. 1996) though occasional 
detections have been made (Latham et al. 1981, Shakhovskoy et al. 1992).
In 1997, at a comparable brightness level, no significant polarization 
is detected in the same R and I bands, with mean values respectively of
(0.8$\pm$1.9)\% and (0.7$\pm$1.8)\%  for July 1 and 
(1.9$\pm$4.2)\% and (0.5$\pm$2.3)\%  for July 30, 
indicating that the cyclotron emission has
become negligible (Shakhovskoy et al. 1992).

\begin{figure*}
\resizebox{\hsize}{!}{\includegraphics{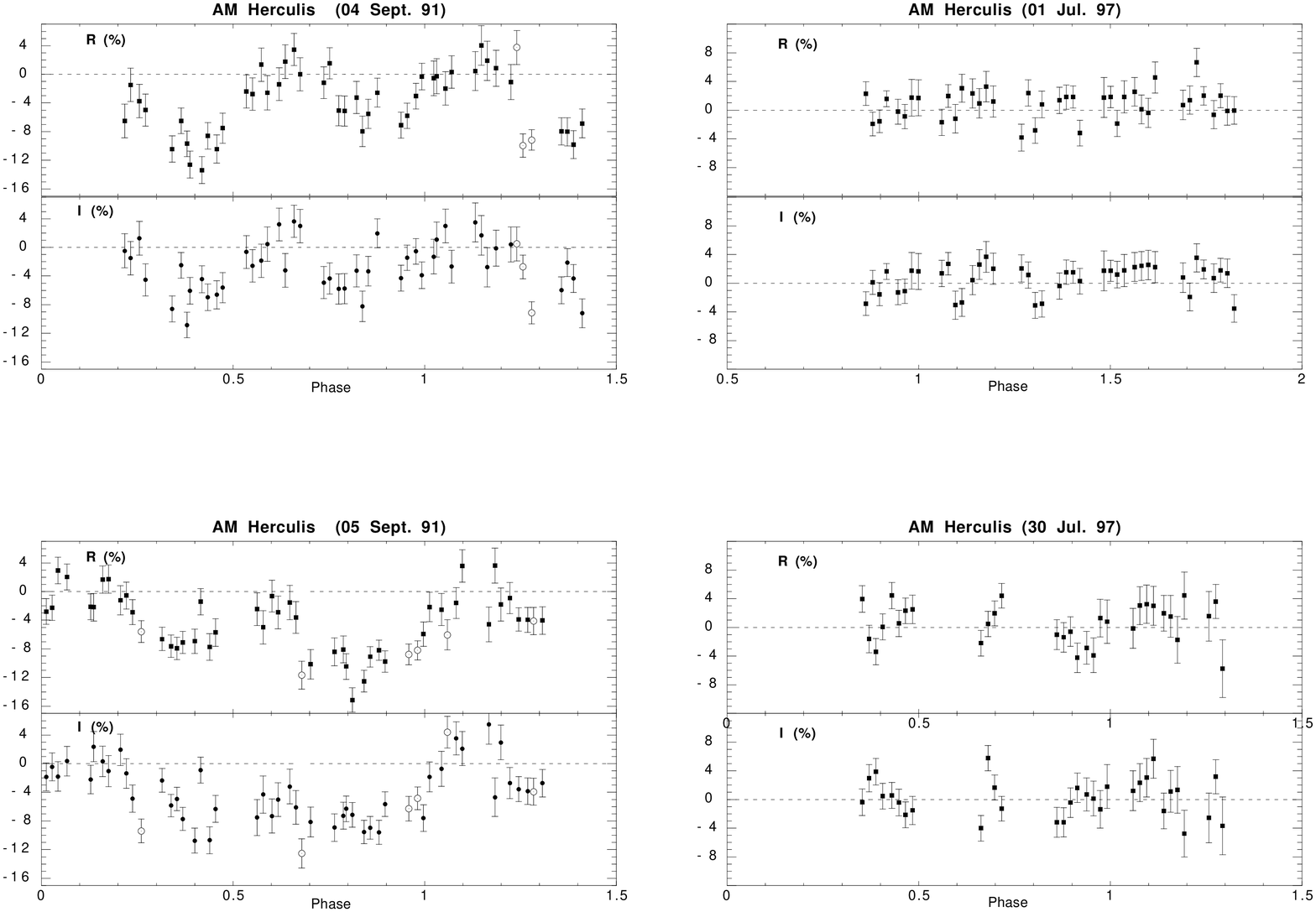}}
%\picplace{12.0 cm}
%psfig{figure=AMfig1a.eps,width=18.cm}
%\scaledpicture 132mm by 99mm (Figure1 scaled 700)
\caption[ ]{UBVRI circular polarization light curves in \am low states
(left July 04 (top) and 05 (bottom), right July 01 (top) and 30 (bottom)). 
Open symbols in 1991 correspond to the flares marked in Figure 1}
\end{figure*}

The peculiar flaring variability of the source seen in 1991 has been 
investigated by computing the characteristics of the flares marked in Figure 1. 
Figure 4a shows the colour-colour diagram of the AM Her system during 
the quiescent state (outside flares) and at the peak of the flares (with the size 
of symbols indicative of the flare intensity).
For comparison, the colours of the very 
large blue flare observed by Shakhovskoy et al. (1993) during a 1992 low state 
are also shown. The AM Her colours during quiescence are remarkably similar in
the different epochs. The flares are clearly distributed into three different 
categories: moderate flares (F3,F5,F6) during which the system stays approximately
of the same colour as in the  quiescent state or slightly bluer, intense flares 
(F1,F2,F4) during which the colours change to strongly red  
and the very intense 1992 flare clearly peaking in the blue.

The energy distribution of the flares has been  
evaluated by computing in the different bands the difference between the magnitude at peak 
and the local quiescent magnitude, estimated from polynomial fits through
the light curve excluding flares.
Magnitude excess due to the flares is reported in Table 3 and the corresponding
spectra of selected flares are given in Figure 4b.
The slope of the best linear fit to the (log\,F$_{\nu}$-log\,$\nu$) distribution 
is also given in Table 3. 
As indicated by the slope, the strongest flares appear 
clearly red, peaking around R/I bands with a slope $\sim\nu^{-(2-3)}$, while the 
less intense ones are bluer $\sim\nu^{-(1)}$.
We note that the polarization during the largest flares is significant in both R and I bands
(see open symbols in Fig. 3a and b).
The properties of the flares are further discussed in Section 4.

\begin{figure}
\resizebox{\hsize}{!}{\includegraphics{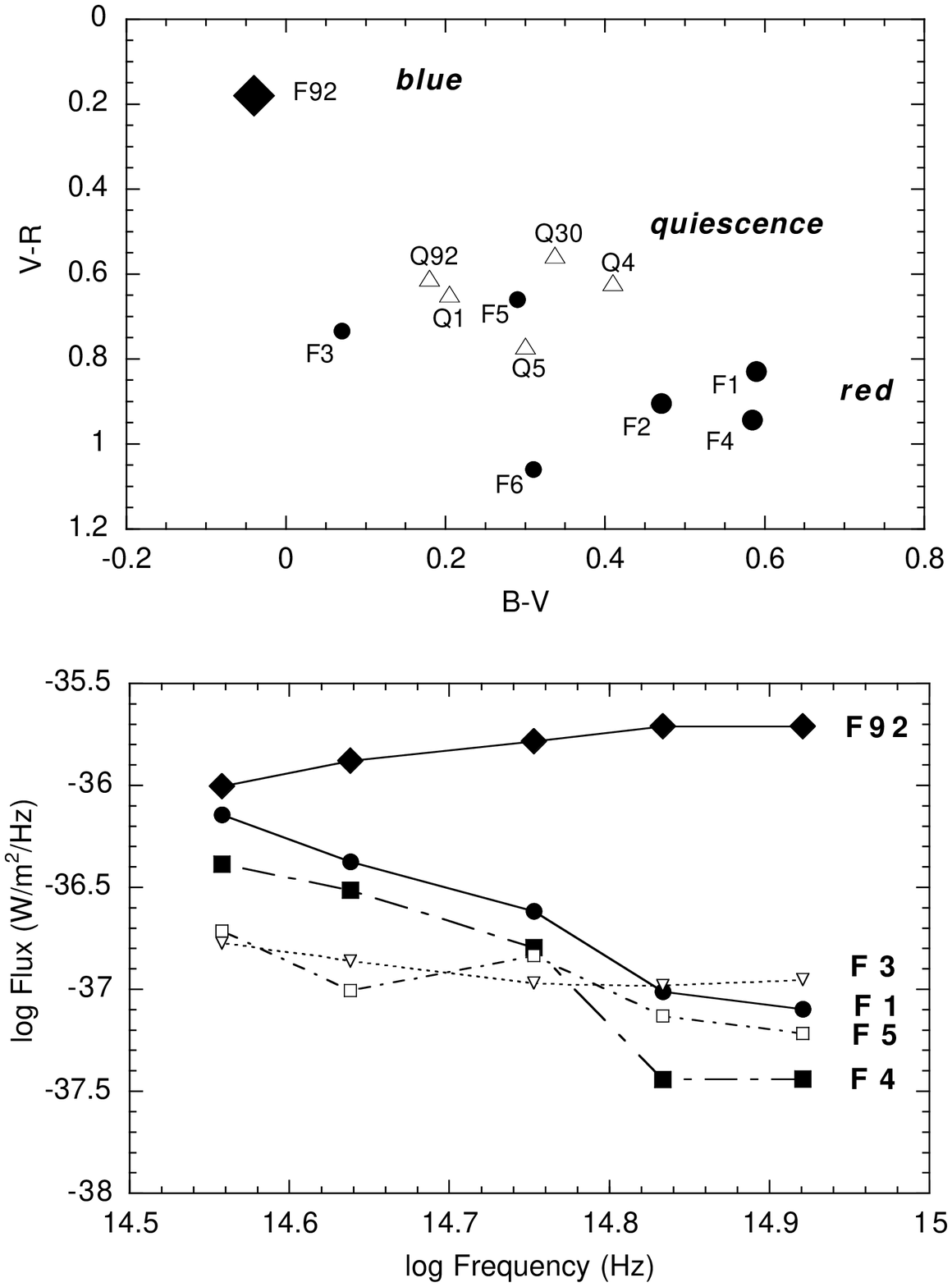}}
%\picplace{12.0 cm}
%\scaledpicture 132mm by 99mm (Figure2 scaled 700)
\caption[ ]{a (top)- (V-R/B-V) colour-colour diagram of AM Her during the flares (F)
and during the quiescent (Q) low states. The size of the flare symbols are indicative
of their intensities.\\ 
b(bottom)- Energy distribution (logF$_{\nu}$-log\,$\nu$) of the light excess due to the flares.}
\end{figure}

\begin{table*}
\caption{Magnitude excess due the flares with respect to the local quiescence level.}
\medskip
\begin{tabular}{rrrrrrrr}
\multicolumn{1}{c}{\em Time(HJD)} & \multicolumn{1}{c}{\em Flare} &
  \multicolumn{1}{c}{\em U} & \multicolumn{1}{c}{\em B} &
  \multicolumn{1}{c}{\em V} & \multicolumn{1}{c}{\em R} & 
  \multicolumn{1}{c}{\em I} & \multicolumn{1}{c}{\em $\alpha$(F$_{\nu}$$\sim\nu^{\alpha}$)}\\
  \hline
8864.3612  &  F92  & -2.54   & -2.36   & -2.14  & -1.71 & -0.84 & +0.82 (0.14)\\
8504.3862  &  F1   & -0.32   & -0.37   & -0.67  & -0.77 & -0.63 & -2.75 (0.26)\\
8505.2875  &  F2   & -0.13   & -0.13   & -0.24  & -0.31 & -0.30 & -3.09 (0.30)\\
8505.3400  &  F3   & -0.37   & -0.35   & -0.38  & -0.33 & -0.20 & -0.53 (0.17)\\
8505.3799  &  F4   & -0.15   & -0.13   & -0.43  & -0.47 & -0.35 & -3.29 (0.59)\\
8505.3903  &  F5   & -0.26   & -0.28   & -0.53  & -0.23 & -0.24 & -1.19 (0.45)\\
8505.4135  &  F6   & -0.20   & -0.25   & -0.27  & -0.44 & -0.42 & -2.79 (0.36)\\
\hline
\multicolumn{5}{l} {Error bars are given in parentheses}\\
\end{tabular}
\end{table*}

Quasi-periodic oscillations (QPO) have been searched using
(0.1s) resolution photometric data obtained in 1991. 
While no significant QPOs have been found on 1991 September 4,
QPOs are clearly detected on 1991 September 5, with an amplitude 8.6\% and 
a period 6.6 min (397s)  during 30 minutes covering a
(0.97-1.14) phase interval. No QPOs were observed in 1997.\\
 
\begin{figure}
\resizebox{\hsize}{!}{\includegraphics{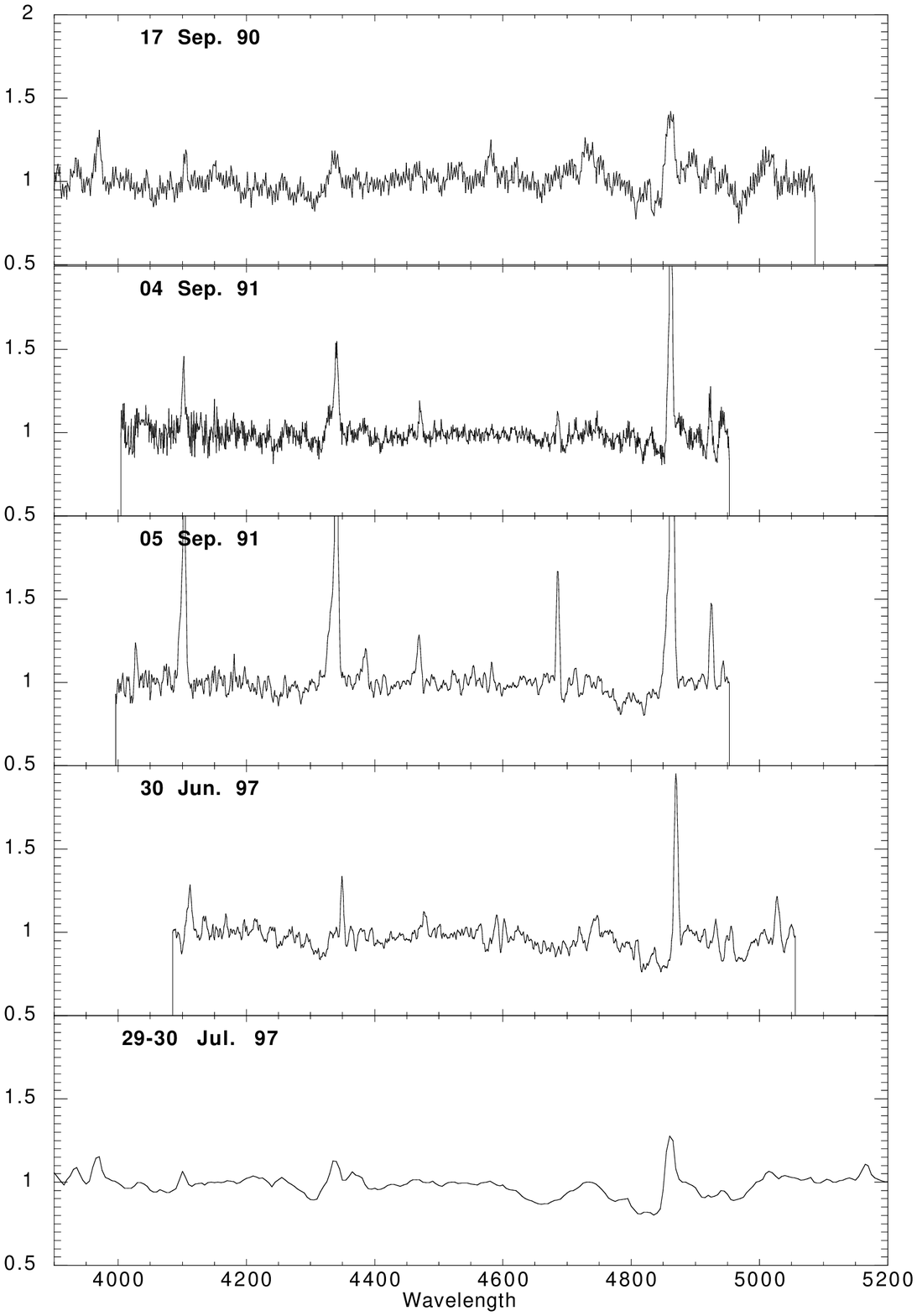}}
%\picplace{12.0 cm}
%\scaledpicture 132mm by 99mm (Figure2 scaled 700)
\caption[ ]{Mean low state spectra for the different epochs. Note the strong 
increase in the emission lines from 1991 Sept. 4 to 5 and Zeeman absorptions at
Balmer lines visible in all spectra.}
\end{figure}

\subsection{Spectroscopy}  
The spectra obtained during the 1990, 1991 and 1997 low states 
have been averaged to produce a mean normalized spectrum 
representative of each different epoch (Figure 5). 
The spectra were reduced by standard procedure using MIDAS-ESO package and
fluxes have been normalized by dividing by a continuum fitted through
selected points free of line features.
All spectra show clear evidence of the Balmer lines in emission with
their associated Zeeman components in absorption, similar to what previously
observed (Latham et al. 1981, Schmidt et al. 1981, Young et al. 1981, 
Silber et al. 1996).  
The equivalent widths of the emission lines appear to vary in the different 
spectra(see Table 4). The HeII 4686 line is only detectable in 1991. 
At this epoch,  the spectrum changes drastically
from one day to the next, with a sudden appearance of this high excitation line, 
and an equivalent width changing from $(0.4\pm0.5)$\AA\, to $(3.6\pm0.5)$\AA. 
On 1991 Sept. 5, though the system is at the same mean low level (m$_{V}$ = 15.1)
and the Zeeman absorption lines, typical of low state, 
are still clearly present, the emission lines are strong with \Hbet and \Hgam
EW of 19\AA\, and 14\AA\, respectively, within a factor 1.5-2 of the high state 
values. This suggests a significant residual accretion.\\
The location and depth of the Zeeman absorption features are best shown in Figure 6, 
where the mean of all low and medium resolution spectra are displayed on an 
extended scale, with the emission part being cut. 
The two sets of data appear very similar with, at higher resolution, the evidence 
of clear and rather sharply defined absorptions around 4080, 4300, 4650 and
4820\AA. To identify these features, synthetic idealized spectra have been
constructed using the Zeeman wavelengths and oscillator strengths tabulated by Kemic (1974)
for different magnetic fields in the range of 10-30MG as expected for AM Her.
The intensities of the lines were taken to be proportional to the oscillator 
strengths 
in the Milne-Eddington approximation (see Latter et al. 1987)
and the spectra were 
interpolated with respect to the field strength to provide a complete grid of
comparison spectra.
The computed "theoretical" spectra were further convolved with an instrumental
response corresponding to a spectral resolution  of 2\AA.\\
The inspection of the synthetic spectra reveals that the most significant 
features of Fig. 6 correspond to the $\sigma$ and $\pi$ components of the \Hbet
and \Hgam hydrogen lines. In the range of considered B fields,
these lines clearly distributed into "stationary" lines,
only weakly variable in position, such as \Hbet $\sigma^{+}$ and $\pi$ and "non 
stationary lines" strongly or moderately variable in position such as 
\Hbet $\sigma^{-}$ and \Hgam $\pi$ (Angel 1978).
The non stationary lines allow a precise determination of the magnetic field
strength. In Fig. 6 is shown  the B=12.5MG synthetic spectrum which is the
best description of the data. Most features are accurately reproduced such as the
split components of \Hbet and \Hgam. The accuracy in the B determination is 
typically $\pm0.5$MG based on a precision better than $\sim$ 10\AA\ in the position of 
the fast varying \Hbet $\sigma^{-}$ feature.\\
 More careful inspection reveals two additional features at
$\sim$ 4400\AA\, and $\sim$ 4775\AA\, that are not
reproduced by the synthetic spectrum. 
Interestingly enough, the $\sim$ 4775\AA\, feature, which is clearly visible 
in Fig. 6 as a left shoulder to the \Hbet $\pi$ feature, can be reproduced
by a \Hbet $\pi$ component from a significant higher field ($\sim$ 17MG). 
The corresponding 
\Hbet $\sigma^{-}$ will then be shifted to $\sim$ 4560\AA\, where a small
feature is indeed present in the spectrum so we cannot exclude the possible 
superposition of this higher field.
A similar additional higher field component has already been reported by Latham et al. (1981).
We also investigated the possible contributions 
from helium lines. For fields in the polar range, Zeeman helium stationary components 
are expected from the HeI 4471\AA\, line at 4320, 4420 and 4530\AA\,,
not in accordance with the observed features. \\
We looked for possible variations of the main Zeeman features 
with orbital phase using our longest set of data of 1997 July 29 at low
resolution. The spectrum appears remarkably stable in phase with only possible
minor variations in intensities around the \Hbet $\sigma^{-}$ line.

\begin{table}
\caption{Emission line equivalent widths (in \AA).}
\medskip
\begin{tabular}{lllll}
\multicolumn{1}{c}{\em Date} & \multicolumn{1}{c}{\em \Halp} &
  \multicolumn{1}{c}{\em \Hbet} & \multicolumn{1}{c}{\em \Hgam} &
  \multicolumn{1}{c}{\em \HeII}    \\  \hline
90 Sep 17  &  -    & 6.0(5)    & 2.0(5)   & -   \\
91 Sep 04  &  -    & 8.7(5)    & 4.8(6)   & 0.4 (5) \\
91 Sep 05  &  -    & 19.0(8)   & 14.1(3)  & 3.6 (5) \\
97 Jun 30  &  -    & 6.0(6)    & 1.4(3)   & - \\
97 Jul 01  & 32(1) & 4.1(5)    & 1.8(2)   & - \\
97 Jul 29  & 39(1) & 5.6(5)    & 2.8(3)   & - \\
97 Jul 30  & 25(1) & 2.5(6)    & 1.2(3)   & - \\
\hline
\multicolumn{5}{l} {Error bars on last digit are given in parentheses}\\
\end{tabular}
\end{table}

\section{Discussion}
\subsection{The AM Her magnetic field}
The best magnetic field derived from the Zeeman absorption components 
in the AM Her optical spectrum during different episodes of low state in 1990, 1991 
and 1997 is B=(12.5$\pm$0.5)MG, in accordance with what previously reported
but with better accuracy (Latham et al.\,1981, Schmidt et al.\,1981, Young et al.\,1981, 
Silber et al. 1996).
The Zeeman split absorption components in polar low states are 
usually assumed to originate in the hot photosphere of the magnetic white dwarf
whose contribution becomes dominant when the accretion ceases.

The relative good accuracy in the magnetic field determination is achieved by the
stability of "non stationnary" components such as \Hbet $\sigma^{-}$
which are very sensitive to the field strength. Despite the spectra are averaged
over different orbital phase intervals and different epochs, the presence of such
a stable feature is an indication of a rather homogeneous field. 
This situation is surprising since, in a simple dipole model, a range of 2 is expected 
between the polar and equatorial field and a spread of B-values is therefore expected 
for an hemisphere seen at a given inclination (Saffer et al. 1989). 
The observed restricted range would therefore imply a nearly equator-on view with
an inclination close to 90$^{\circ}$ and an observed B field peaking 
at B$_{polar}$/2.

\begin{figure*}
\resizebox{\hsize}{!}{\includegraphics{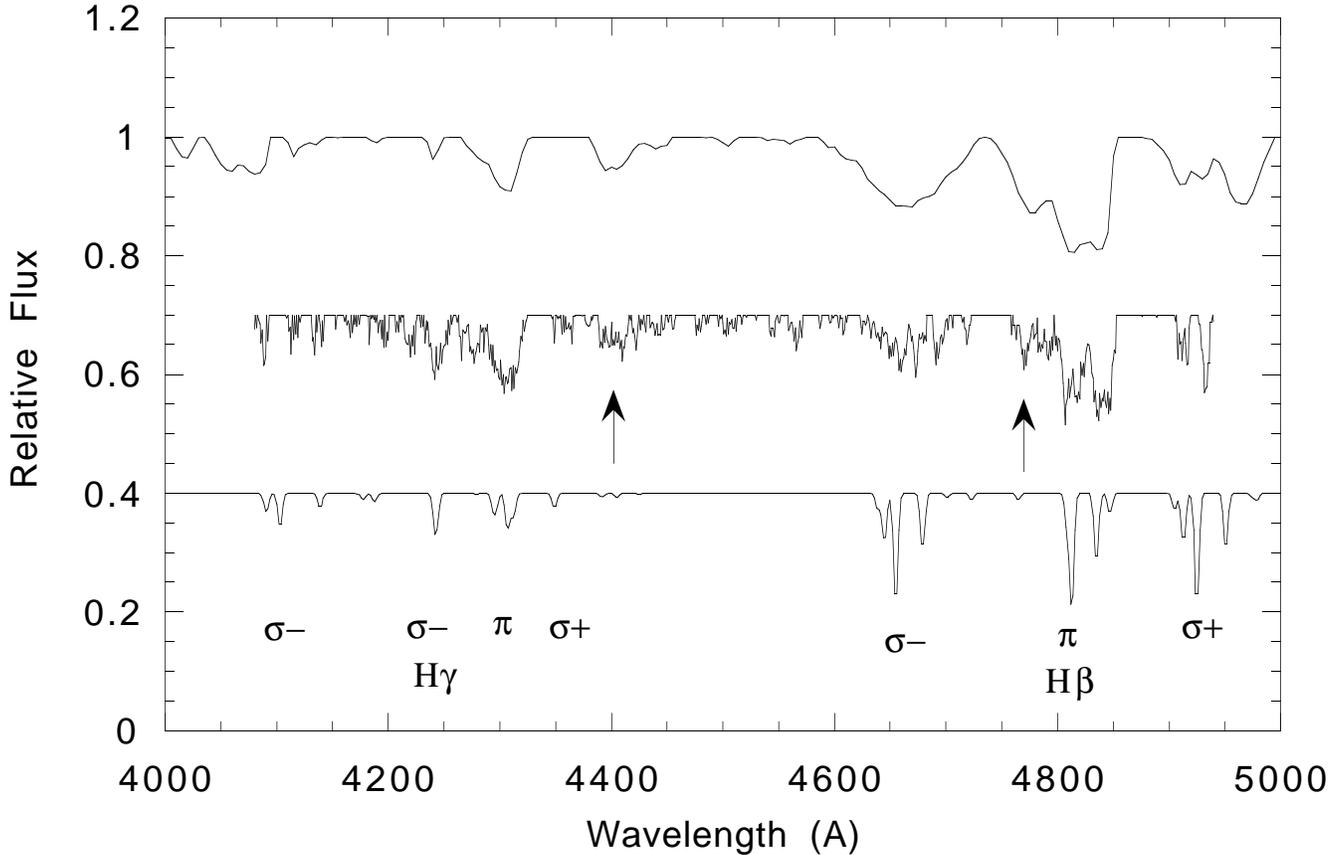}}
%\picplace{6.0 cm}
%\scaledpicture 132mm by 99mm (Figure2 scaled 700)
\caption[ ]{Low state spectra showing clear Zeeman absorptions. The normalized 
continuum has been cut to suppress the emission lines. The low
resolution spectrum (top) corresponds to the 1997 low state only while the 
higher resolution one (middle) is an average of 1990, 1991 and 1997 low states. 
The main Zeeman identifications are shown. The fitted best (B=12.5 MG)
synthetic spectrum is displayed at bottom and reproduces most of the strongest features. 
The arrows indicate unidentified features (see text) }
\end{figure*}

\subsection{Unstable accretion during the low states}

The AM Her optical low states reported in this paper, 
though all showing a similar optical brightness with  m$_v$=(15.1-15.4), 
display different overall characteristics. 
Of particular interest is the behaviour of the source in 1991
when both a significant  polarization and flaring activity is seen 
contrary to 1990 and 1997.
Within the brightness history of the source (Mattei  J., AAVSO, 
private communication), it may be significant that the 
1991 observation is located toward the end 
of a rather slow decline from high to low state, while the 1990 and 
1997 observations are both included in a prolonged long stable low state, 
lasting already for 5 and 2 months respectively.\\
The major flares observed in 1991 are predominantly red in nature
when compared to quiescent state. This and the presence of a significant
polarization point toward residual unstable accretion at that time.
The presence of a low level ($\sim$5min.) temporary QPOs
usually observed during intermediate high states (Bonnet-Bidaud et al. 1992)
and the reappearance of high excitation lines on Sept 5, 1991 
further strengthen this conclusion.

The comparison of this flaring activity with the major flare 
observed in 1992, also during a low state, is interesting.
It has been proposed that the sharp 1992 event is due to a stellar flare
originating from a magnetic active secondary (Shakhovskoy et al. 1993).
The shape as well as the colour changes of the flare were found consistent with
what observed during strong flares from red dwarfs (Beskrovnaya et al. 1996).
The total energy in the flare, though at the very extreme upper end of what usually 
observed in nearby UV Ceti systems, is comparable to the more energetic flares
detected in open clusters (Shakhovskaya 1989).  
However flares of this type are also
fairly repetitive while it has been observed only once in AM Her although 
the source has been intensively monitored. We note however that a
flare at this amplitude may be lost during high states.
We find significant that the maximum magnitude of the flare 
(m$_v$$\sim$12.5) is of the order of the high state level so that this episod
may also be interpreted alternatively as an unstable accretion event.
The very peculiar colours of this flare (see Fig. 4a) may result in
this case from inhomogeneous accretion with blobs buried into the white 
dwarf atmosphere if the density is high enough.\\
Following Frank et al. (1988), we estimate the critical density for
buried shocks by equating the accreted gas ram pressure with the 
atmosphere pressure at an optical depth sufficient for 
efficient reprocessing, giving 

$\rho_{cr}$ = 1.42 $M_{wd}^{2}$. [R$_{wd}^3$.T$_{wd}$.ln(97R$_{wd}$.$\rho_{cr}$)]$^{-1}$

where $\rho_{cr}$ is the critical density in units of 10$^{-6}$ g.cm$^{-3}$
and $M_{wd}$, R$_{wd}$ and T$_{wd}$ the mass, radius and temperature of
the white dwarf in units of M$_{\sun}$, 10$^{9}$cm and 10$^{5}$K
respectively.
Assuming T$_{wd}$$\sim$ 0.2 and $M_{wd}$=0.6 (see Gänsicke et al. 1995) 
with a corresponding R$_{wd}$=0.9 (Nauenberg 1972) gives
a critical density of $\rho_{cr}$ = 0.85 10$^{-6}$g.cm$^{-3}$.
This has to be compared with the mean accreted density during steady
high states $\rho_{high}$ = $\dot{M}$/(v$_{ff}$A) with a typical 
accretion rate $\dot{M}$ 
$\sim 10^{16}$ g.s$^{-1}$ , v$_{ff}$ the
free-fall velocity $\sim 10^{8}$ cm.s$^{-1}$  and A the accreting area
$\sim 10^{16}$ cm$^{2}$ which yields $\rho_{high}$ = 10$^{-8}$g.cm$^{-3}$.\\
To achieve blobby accretion, the density in the flare has then to be
$\sim$10-100 times the steady accretion. This may be easily achieved if, for instance, the
accretion during the temporary event takes place onto a (3-10) reduced 
accretion spot radius due to particular unstable capture conditions by the magnetic 
field at that time. 
The emerging radiation from such a blobby accretion has not yet been 
computed accurately but the radiation is expected to be thermalized inside the 
atmosphere and will be radiated at temperature closer to the white dwarf blackbody 
(Kuijpers \& Pringle 1982). 
We then expect the optical spectrum of the flare to follow roughly a 
($\sim \nu^{+(1-2)}$) dependency, in accordance with the observed colours. 
Such flares should be mostly visible in soft X-rays.\\
On the opposite, the smaller flares observed during low states may correspond to
small accretion events during which the accretion rate only temporarily increased,
leading to transient presence of a standard shock above the white dwarf and 
an associated cyclotron emission. Such small scale flares, as those observed in 1991,
show indeed polarization and red colours  ($\sim \nu^{-(2-3)}$) expected from 
the optically thin part of the cyclotron.
During the typical low states, outside flares, the low density accretion flow may lead
to the absence of a shock and heat the upper atmosphere by Coulomb collisions 
leading to the so called bombardment solution (Woelk \&Beuermann 1992). 
In this last case, the bluer colors of the optical radiation during quiescence can be 
easily explained by the decrease (1991) or negligible (1997) cyclotron emission.
This picture derived from the optical variability study of AM Her is in accordance 
with the conclusions drawn from the low state X-ray characteristics of the source 
(Ramsay et al. 1995).
The reason of the unstable accretion during low states, leading to both large 
accretion event and/or small accretion instability, is still unresolved as well as
the exact mechanism responsible for the low states.

\section{Conclusion}
The study of the temporal and spectral characteristics of AM Her during low states
at different periods allowed different conclusions to be drawn.\\
The Zeeman spectral features shown by the source are surprisingly very stable
although obtained through different parts of the orbital cycle and therefore
with different orientations with respect to the suspected dipole magnetic field.
The magnetic field strength derived from the position of "non stationary" lines 
is (12.5$\pm$0.5)\,MG which could represent an averaged field seen over the white
dwarf surface. Additional features are seen which may originate from an higher field 
region.\\
The temporal optical variability of AM Her during low states is very rich, displaying 
occasional large blue flare events as well as repetitive smaller amplitude 
red flares. It is shown that the characteristics of all these flares can be explained by
accretion events of different amplitudes. The large and unique event observed in 1992,
though consistent with red dwarf flares, can also 
be tentatively explained by a large increase of the accretion rate 
coupled to a reduced accretion area which can lead to a sufficient increase in the density
to produce a buried shock unstable accretion. The more frequent smaller amplitude flares 
are interpreted instead as small variable increases of the accretion rate.\\

{}

\end{document}